\documentclass[fleqn,usenatbib]{mnras}
\usepackage{graphicx}
\usepackage{amsmath}
\usepackage{amssymb}
\usepackage{multicol}
\usepackage{bm}
\usepackage{pdflscape}
\usepackage[T1]{fontenc}
\usepackage{ae,aecompl}
\usepackage{xcolor}

\title[Hybrid cooling Cas A NS]{Hybrid cooling of the Cassiopeia A neutron star}
\author[Lev B. Leinson]{Lev B. Leinson$^{1}$\thanks{E-mail: leinson@yandex.ru} \\
$^{1}$Pushkov Institute of Terrestrial Magnetism, Ionosphere and Radiowave
Propagation of the Russian Academy of Science (IZMIRAN), \\
108840 Troitsk, Moscow, Russia}
\date{Accepted XXX. Received YYY; in original form ZZZ}
 \pubyear{2018}

\begin{document}
\label{firstpage}
\pagerange{\pageref{firstpage}--\pageref{lastpage}}
\maketitle 

\begin{abstract} 
The observed rapid cooling of the neutron star Cassiopeia A is usually interpreted as being caused by transitions of neutrons and protons in the star's core from the normal state to the superfluid and superconducting state. However, this so-called "minimal" cooling paradigm faces the problem of numerically simulating the observed anomalously fast drop in the neutron star surface temperature using theoretical neutrino energy losses from superfluid neutrons. As a solution to this problem, I propose a somewhat more complex cooling model, in which, in addition to superfluid neutrons, direct Urca processes from a very small central part of the neutron star core are also involved. Numerical simulations of the cooling trajectory in this scenario show excellent agreement with observations of the Cassiopeia A neutron star. 
The proposed cooling scenario unambiguously relates the used equation of state and the mass of the neutron star. For a neutron star constructed according to BSk25 equation of state, the most appropriate are the mass $M=1.62M_{\sun}$ and the radius $R=12.36$ km. If BSk24 equation of state is used, then the most suitable solution is $M=1.60M_{\sun}$ and $R=12.55$ km. 
\end{abstract}

\section{Introduction}
\label{sec:intro}

The neutron star (NS) in the supernova remnant Cassiopeia A (Cas A),
discovered by the Chandra satellite in 1999 \citep{Hughes:1999ph}, is of exceptional
interest to astrophysicists. At a sufficiently low surface magnetic field of 
$10^{10}\div 10^{11}$ G, the X-ray radiation of this hot isolated superdense
object can be unambiguously associated with the temperature of its surface,
neglecting the non thermal X-ray radiation of the magnetosphere. In this
regard, the study of the thermal evolution of this young \citep[$\approx 340$ yr
old,][]{Fesen:2006zma} neutron star in X-rays is of great importance for a better
understanding of the evolution of such superdense objects and makes it
possible to study their composition and structure 
\citep[see, e.g.][]{Page:2004fy,Page:2009fu,Yakovlev:2004iq}.

About ten years ago, \citet{Ho:2009mm} and \citet{Heinke:2010cr} analyzed a decade of Chandra observations of the Cas A supernova remnant and reported a sustained anomalous drop in the star surface temperature $T_{e}$, a phenomenon that had never been observed for any isolated neutron star. 
Despite significant corrections to the observational data over next 10 years, the decrease in the temperature of the neutron star is still anomalously fast. 
Current analysis \citep{Posselt:2018xaf,Wijngaarden:2019tht,Ho:2021hwy}
gives upper limits corresponding to a 3.3\% or 2.4\% decrease in temperature
over 10 years depending on values of the absorbing hydrogen column density.
Such a rapid drop in surface temperature contradicts the standard neutrino cooling scenario based on efficient modified Urca processes. If the NS in Cas A were to undergo standard cooling, then the drop in its surface temperature over 10 years would be $0.2\%-0.3\%$ \citep{Yakovlev:2000jp,Page:2005fq}.

To explain these observations, various exotic energy loss mechanisms have been 
proposed using nonstandard assumptions about the physics and evolution of NSs, 
including softened pion modes \citep{Blaschke:2011gc}, quarks 
\citep{Sedrakian:2013pva,Noda:2011ag} and cooling after heating process in 
r-mode \cite{Yang:2011yg} or turbulent magnetic field 
\citep{Bonanno:2013oua}. Other scenarios have been proposed, suggesting the so-called "enhanced" cooling due to direct Urca processes \citep{Negreiros:2011ak,Taranto:2015ubs} or additional emission of axions.  
\citep{Leinson:2014ioa,Leinson:2021ety,Hamaguchi:2018oqw}. 
Except in the case of additional energy losses because of the axion emission, the rapid cooling in these scenarios occurs from the birth of a NS, so the current temperature must be much lower than actually measured. 
The rapid decrease, but relatively high surface temperature (about $2\times 10^{6}$ K) indicates that the cooling was slow at first, but then accelerated significantly, which requires a sharp change in the properties of neutrino emission from NS. 
\citep{Page:2010aw,Shternin:2010qi}.

With this in mind, it has been suggested that the observed rapid temperature drop can be naturally explained in terms of the minimum cooling paradigm 
\citep{Page:2004fy,Page:2009fu}, where the rapid cooling of a neutron
star is caused by the neutron superfluidity in the core. This scenario assumes that neutrons have recently become superfluid (in the $^{3}$P$_{2}$ state) in the NS core, causing a huge flux of neutrinos as a result of the Cooper pairs breaking and formation (PBF) in thermal equilibrium 
\citep{Page:2010aw,Shternin:2010qi}, and protons already in the superconducting singlet state 
$^{1}$S$_{0}$ with a higher critical temperature suppress modified Urca processes. 

The PBF emission of neutrinos occurs mainly owing to spin fluctuations of
neutrons, since in the considered long-wavelength limit, the emission of
neutrinos in the vector channel is completely suppressed \citep{Leinson:2006gf,Steiner:2008qz}.
The correct form of the PBF neutrino emissivity of the $^{3}$P$_{2}$
superfluid neutrons, as derived in \citep{Leinson:2009nu}
\citep[for recent review see][]{Leinson:2017dlo}, reads:  
\begin{equation}
Q_{\bar{\nu}\nu }^{\mathrm{PBF}}\simeq \frac{2}{15\pi ^{5}}G_{F}^{2}C_{%
\mathsf{A}}^{2}p_{Fn}m_{n}^{\ast }\mathcal{N}_{\nu }T^{7}F_{4}\left(
T/T_{cn}\right) ~,  \label{Qnu}
\end{equation}%
where $G_{F}=1.166\times 10^{-5}$ GeV$^{-2}$ is the Fermi coupling constant, 
$C_{\mathsf{A}}=1.26$ is the axial-vector coupling constant of neutrons, $%
p_{Fn}$ is the Fermi momentum of neutrons, $m_{n}^{\ast }\equiv
p_{Fn}/V_{Fn} $ is the neutron effective mass, $\mathcal{N}_{\nu }=3$ is
the number of neutrino flavors, and $T_{cn}$ is the critical  temperature for neutron pairing; 
the function $F_{l}$ is given by 
\begin{equation}
F_{l}\left( T/T_{c}\right) =\int \frac{d\mathbf{n}}{4\pi }\frac{\Delta _{%
\mathbf{n}}^{2}}{T^{2}}\int_{0}^{\infty }dx\frac{z^{l}}{\left( \exp
z+1\right) ^{2}},  \label{F0}
\end{equation}%
with $\mathbf{n}=\mathbf{p}/p$ ($\mathbf{p}$ is a quasiparticle momentum) and 
\begin{equation}
z=\sqrt{x^{2}+\Delta _{\mathbf{n}}^{2}/T^{2}}.  \label{z}
\end{equation}%
The superfluid energy
gap\footnote{\label{fn4}The definition of the gap amplitude in Eq.
(\ref{Dn}) matches what is implemented in the public NSCool code by 
\citet{Page16} 
and differs from
the gap definition used in refs. \citep{Leinson:2009nu,Leinson:2017dlo} by 
$1/\surd 2$ times.}
\begin{equation}
\Delta _{\mathbf{n}}\left( \theta ,T\right) =\ \Delta \left( T\right) \sqrt{%
1+3\cos ^{2}\theta },  \label{Dn}
\end{equation}%
is anisotropic. It depends on polar angle $\theta $ of the quasiparticle
momentum and temperature $T$.

In standard physical units Eq. (\ref{Qnu}) takes the form\footnote{%
Notice, the neutrino emissivity, as indicated in Eq. (\ref{Qnu}), is 4 times
less than that implemented in the NSCool code.} 
\begin{eqnarray}
Q_{n\nu }^{\mathrm{PBF}} &=&\frac{4G_{F}^{2}p_{Fn}m_{n}^{\ast }}{15\pi
^{5}\hbar ^{10}c^{6}}\left( k_{B}T\right) ^{7}\mathcal{N}_{\nu }\frac{C_{%
\mathsf{A}}^{2}}{2}F_{4}\left( \frac{T}{T_{cn}}\right)  \label{QNu} \\
&=&1.170\,\times 10^{21}\frac{m_{n}^{\ast }}{m_{n}}\frac{p_{Fn}}{m_{n}c}%
T_{9}^{7}\mathcal{N}_{\nu }\frac{C_{\mathsf{A}}^{2}}{2}F_{4}\left( \frac{T}{%
T_{cn}}\right) ~\ \frac{\mathrm{erg}}{\mathrm{cm}^{3}\mathrm{s}}~ \nonumber
\end{eqnarray}%
with $T_{9}=T/10^{9}\mathrm{K}$ and $m_{n}$ being the bare neutron mass.

According to the minimal cooling scenario neutrino energy losses from neutron
pairing in the NS core are responsible for the observed rapid temperature
drop. However, numerical simulations 
\citep{Shternin:2010qi,Shternin:2014fea,Potekhin:2017ufy} have shown that PBF processes in the neutron spin-triplet condensate are not efficient enough to reproduce the observed rapid cooling of Cas A NS reported in \citep{Ho:2009mm,Heinke:2010cr}, which makes it difficult to successfully explain the observational data
\citep{Shternin:2010qi,Shternin:2014fea,Potekhin:2017ufy,Shternin:2021fpt}.

To adapt the simulation result in the minimum cooling scenario with the observational data, the authors increased the PBF neutrino emissivity 
several times by multiplying the expression (\ref{QNu}) by the scaling 
factor \citep{Shternin:2010qi,Elshamouty:2013nfa,Shternin:2021fpt}. This 
approach is questionable because the energy losses, as indicated in Eq. 
(\ref{QNu}), were calculated microscopically, taking into account 
many-particle effects with the same accuracy as other neutrino processes 
involved in the cooling scenario. In particular, this expression involves 
anomalous weak interactions (existing only in superfluid Fermi liquids), 
which significantly reduce the emission of PBF neutrinos. Note that a similar result was obtained in \citep{Kolomeitsev:2008mc} for the PBF neutrinos emitted through the axial channel in the case of neutron spin-singlet pairing. 
It was also demonstrated there \citep[see also][]{2013PhRvC..87b5501L} 
that the polarization effects in nucleon matter 
can only slightly reduce the emission of PBF neutrinos. Thus, the energy 
losses indicated in the Eq. (\ref{QNu}) are quite accurate and cannot be 
arbitrarily enlarged by a scale factor. 

A natural question arises: why the existing theoretical models of the Cas A NS
cooling disagree with observational data? As will be shown below, 
this ten-year-old problem can be easily solved simply by going beyond the 
minimal cooling paradigm and choosing the NS mass so that the direct Urca 
processes are activated in a very small central part of the NS core. The observed anomalous decrease in the NS surface temperature in this case is provided by joint losses of neutrinos from direct Urca reactions and the neutrino emission from PBF processes dominating in the current period. Since this scenario combines "minimal" and "enhanced" cooling, I call it "hybrid" cooling.

In Section \ref{sec:model}, I briefly describe the NS model, including the EOS and superfluid 
and superconducting gaps. In Section \ref{sec:sim} I present the simulation results. 
Finally, I summarize and discuss my findings in Section \ref{sec:disc}.

\section{Neutron star model}
\label{sec:model}

The exact mass of Cas A NS is not directly observed. The measured energy
spectra of the X-ray flux mainly depend on the brightness and gravitational
redshift, which are functions of the NS mass $M$ and radius $R$. Fitting of
the observation\ data with theoretical models showed %
\citep{Elshamouty:2013nfa} that the most suitable mass is $M\simeq
1.62M_{\odot }$, later a wider mass range was proposed, $M\simeq \left(
1.6\div 1.65\right) M_{\odot }$, depending on the accepted equation of state
(EOS) \citep{Heinke:2010cr,Ho:2014pta,Ho:2021hwy,Shternin:2021fpt}. The Cas
A NS model, considered below, uses the most modern EOSs based on the
Brussels-Skyrme nucleon interaction functional, BSk24, BSk25 %
\citep{Pearson:2018tkr}, which have been precisely fitted to almost all
available data on atomic mass and constrained to fit, up to the densities
prevailing in neutron-star cores. Note that the most suitable mass of the
Cas A NS exceeds the fast direct Urca cooling threshold $M_{dU}$ for both
the EOSs: $M_{dU}=1.595M_{\odot }$ and $M_{dU}=1.612M_{\odot }$ for BSk24
and BSk25, respectively. Thus, direct Urca processes are involved in the 
hybrid cooling of Cas A NS. 

Superfluidity of neutrons and superconductivity of protons still play an
important role in the hybrid scenario of NS cooling, partially suppressing
the heat capacity and standard mechanisms of neutrino emission of nucleons.
The most important role is assigned to the enhanced PBF emission of
neutrinos owing to Cooper pairing of neutrons, when the temperature drops just
below the critical value $T_{cn}$ in the NS core. 

To simulate Cas A NS cooling, I use the widely used CCDK model %
\citep{Elgaroy:1996mx} for the $^{1}$S$_{0}$ proton gap, which can create
NSs with a completely superconducting core of protons \citep{Ho:2014pta}. As
was found in \citep{Page:2010aw,Shternin:2010qi}, the critical temperature
for the superfluid transition of neutrons as a function of density should
have a broad peak with a maximum $T_{cn}(\rho )\approx (5-8)\times 10^{8}$
K. Therefore, the TToa \citep{Takatsuka:2004zq} model is used for the $^{3}$P%
$_{2}$ neutron gap in the NS core. For singlet pairing of neutrons  the
"SFB" model \citep{Schwenk:2002fq} is chosen. The choice of other models for
the singlet pairing of neutrons slightly affects the result, since the $^{1}$%
S$_{0}$ pairing of neutrons occurs only in the inner NS crust.  

By solving the standard Tolman-Oppenheimer-Volkoff relativistic equations of 
stellar structure \citep[see, e.g.][]{Shapiro} supplemented by the EOS yields 
the NS mass and radius against central density. 
The evolution of the interior temperature of an isolated NS is determined by
the relativistic equations of energy balance and heat flux 
\citep[see, e.g.][]{2004ARA&A..42..169Y}. 
Cas A NS cooling is simulated using 
the publicly available NSCool code \citep{Page16}, which, along with 
relativistic gravity, 
includes all the corresponding neutrino reactions, heat capacity, and thermal 
conductivity as functions of temperature, density, and composition of superdense 
matter in chemical equilibrium.  The correct form (\ref{QNu}) of the PBF energy 
losses was additionally incorporated. 
\section{Cooling simulation}
\label{sec:sim}
 First, consider BSk25 EOS, which predicts the density threshold for direct
Urca processes $n_{dU}=0.469$ fm$^{-3}$ \citep{Pearson:2018tkr}. 
It is reasonable to study the NS with a central density around this value.

\begin{figure}
\begin{multicols}{2}
\includegraphics[height=8.6cm,trim=120 0 130 0,clip,width=8.6cm]{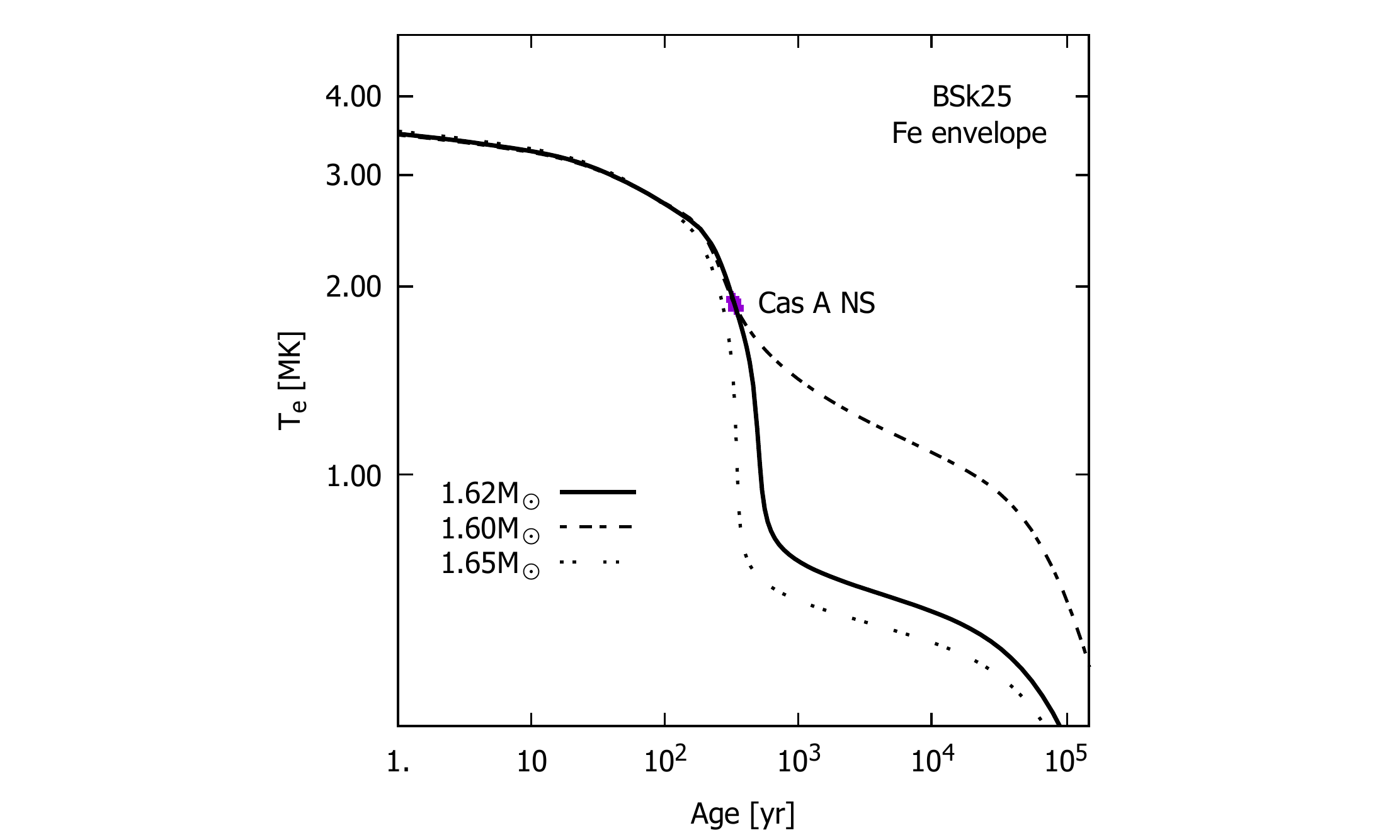}
\includegraphics[height=8.6cm,trim=120 0 130 0,clip,width=8.6cm]{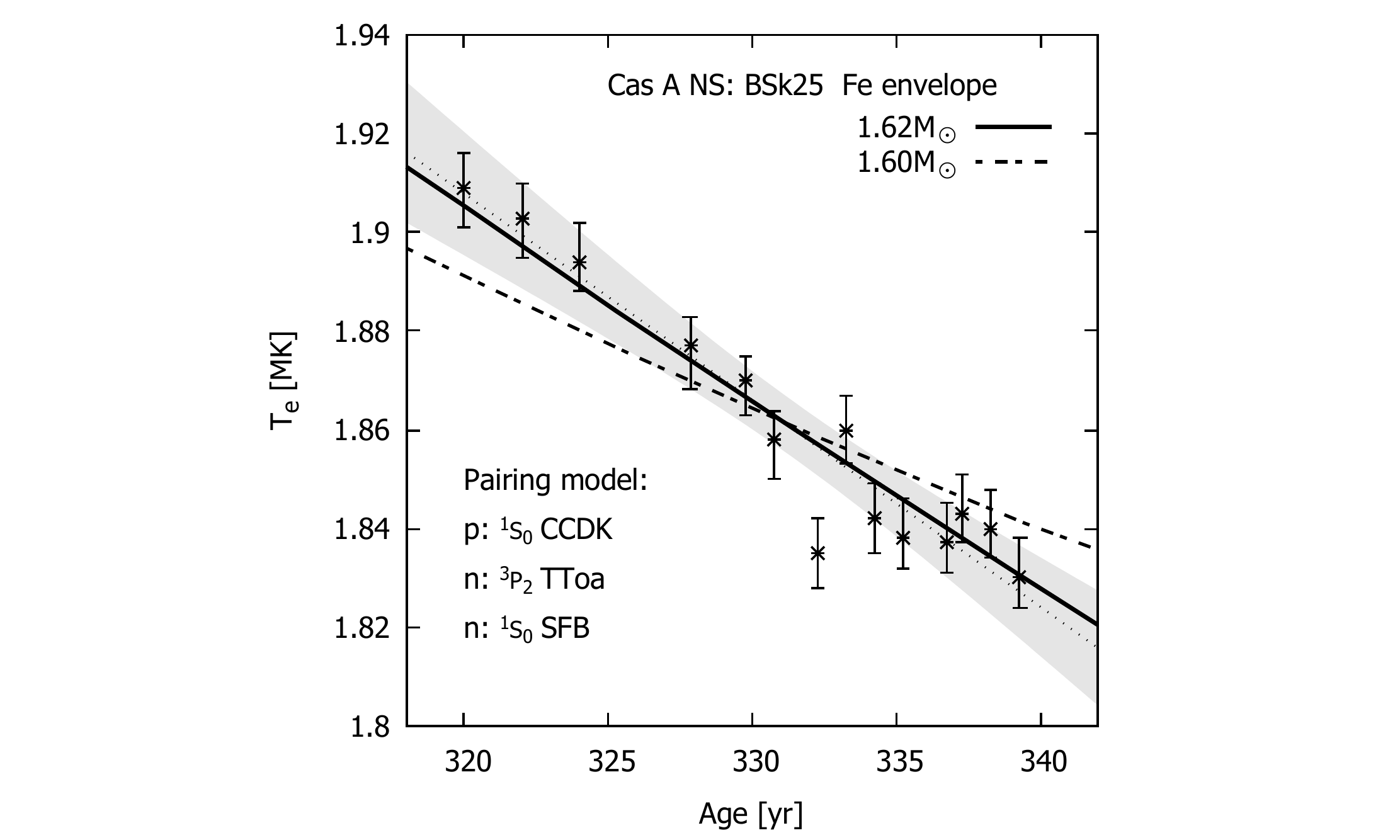}
\end{multicols}
\caption{Dependence of the effective surface temperature $T_e$ without redshift on the age of an isolated neutron star with an iron envelope plotted according to BSk25 EOS. 
Upper panel: the long-term cooling trajectories for 
$M/M_{\odot}=1.6,1.62,1.65$. The spot indicates the current location of Casa NS. 
Lower panel: a comparison of simulated results for 
$M/M_{\odot}=1.6,1.62$ with the  data points of \citet{Ho:2021hwy}. Gray area 
shows the 99\% confidence interval for the Cas A NS surface temperature.}
\label{fig:25}
\end{figure}

The simulation result is demonstrated in Fig.\ref{fig:25}, 
which shows the cooling curves for neutron stars with masses 
$1.60M_{\odot }$, $1.62M_{\odot }$ and $1.65M_{\odot }$. 
The upper panel shows the temporal behavior of the NS surface temperature 
(without gravitational redshift) on a small scale. It can be seen that, in the case of 
$ 1.65M_{\odot} $, NS cooling occurs too quickly, which is caused by the 
participation of a large fraction of nucleon matter in direct Urca reactions. 
In the case of $ 1.60M _{\odot} $ and for $ 1.62M_{\odot} $, the cooling 
trajectories pass through the current Cas A NS location, but they have a 
different slope at this point. 

This is clearly seen in the lower panel, where these cooling trajectories are 
plotted on a large scale.
The asterisks with error bars are the surface 
temperature of a neutron star in a supernova remnant Cassiopeia A, measured 
from the \emph{Chandra} ACIS-S Graded spectra over the past 18 years, as reported in 
\citep{Ho:2021hwy}. (The surface temperature data were recalculated in accordance 
with the mass $M=1.62M_{\odot}$ and radius $R=12.36$ km of the simulated NS, 
assuming that $T_e^2 R = const$ at a fixed distance to the NS.) 
The thin dotted line represents the standard 
linear least squares fit showing the annual average rate of temperature change, 
and the gray area shows the 99\% confidence interval for the Cas A NS surface 
temperature.\footnote{The 99\% confidence interval is built on the assumption 
of a linear temperature versus time, which approximately takes place in the 
considered section of the NS cooling curve.} It can be argued that during the 
observation period the effective surface temperature of the NS should be within 
the gray shaded area with a probability of 99\%.   

For $M =1.60M_{\odot}$, the cooling trajectory does not fall within the 
confidence interval of 99\%, demonstrating a stellar cooling rate below the 
observed one. In contrast, the cooling curve corresponding to 
$M=1.62M_{\odot}$ shows excellent agreement with the observational data.  
This result is easy to understand if one remembers that the threshold of direct 
Urca processes for EOS BSk25 is $M_{dU}=1.612M_{\odot}$. 
A NS with a mass $M=1.60M_{\odot}$ is below this threshold and cools through the 
emission of PBF neutrinos from its core according to the minimal cooling scenario.
As discussed in the introduction, this cooling mechanism 
cannot provide energy losses consistent with the observed rate of temperature 
decrease. The NS mass $M=1.62M_{\odot}$ slightly exceeds the threshold, so direct Urca reactions are activated in a small part of the NS core, causing neutrino energy losses in addition to the PBF emission.

It should be noted that the part of nucleon matter where the direct Urca reactions are active has a very small mass $\Delta M_{dU}=6.770\times 10^{-3}M_{\odot}$, and fills a small central volume with a radius $R_{dU}=1.545\times 10^{5}$ m.\footnote{Therefore a very fine 
grid should be used in numerical simulations increasing substantially the 
calculation time.} 
\begin{figure}
\includegraphics[height=8.6cm,trim=130 0 130 0,clip,width=8.6cm]{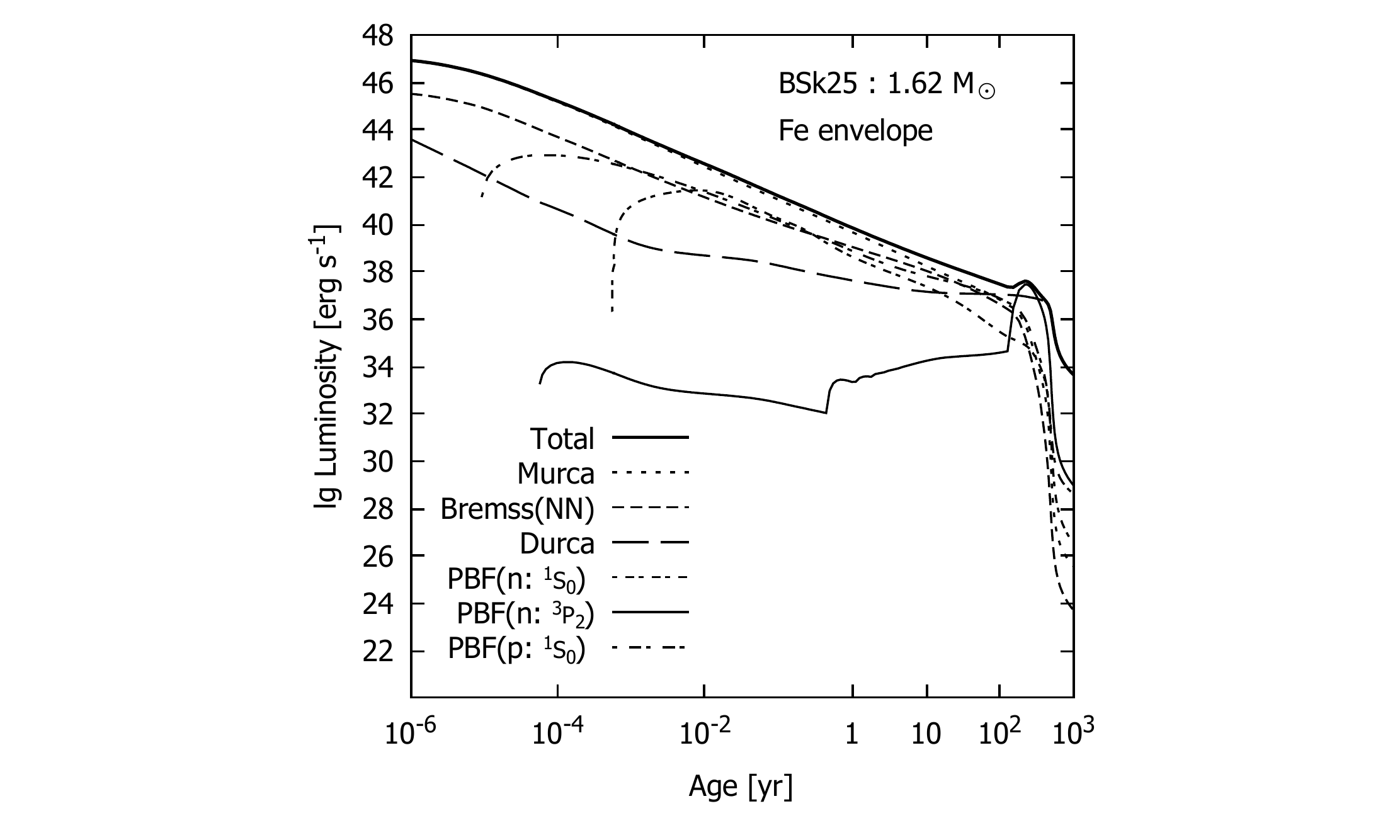}
\caption{The neutrino luminosity from different neutrino sources inside the NS core
versus time.}
\label{fig:lum}
\end{figure} 

The neutrino luminosity from various neutrino sources inside the NS core is 
shown in Fig. \ref{fig:lum} versus time. It can be seen that neutrinos from direct Urca reactions are not dominant in the energy losses, although they act immediately from the moment the neutron star is born. 
When the age of the NS reaches $\sim 300$ years (the current age of the NS Cas A), a burst of PBF neutrinos is observed, which dominates the energy losses and causes the NS to cool rapidly. 
The next in the energy loss intensity during this period are direct Urca reactions, which additionally contribute to faster cooling. 

\begin{figure}
\includegraphics[height=8.6cm,trim=140 0 130 0,clip,width=8.6cm]{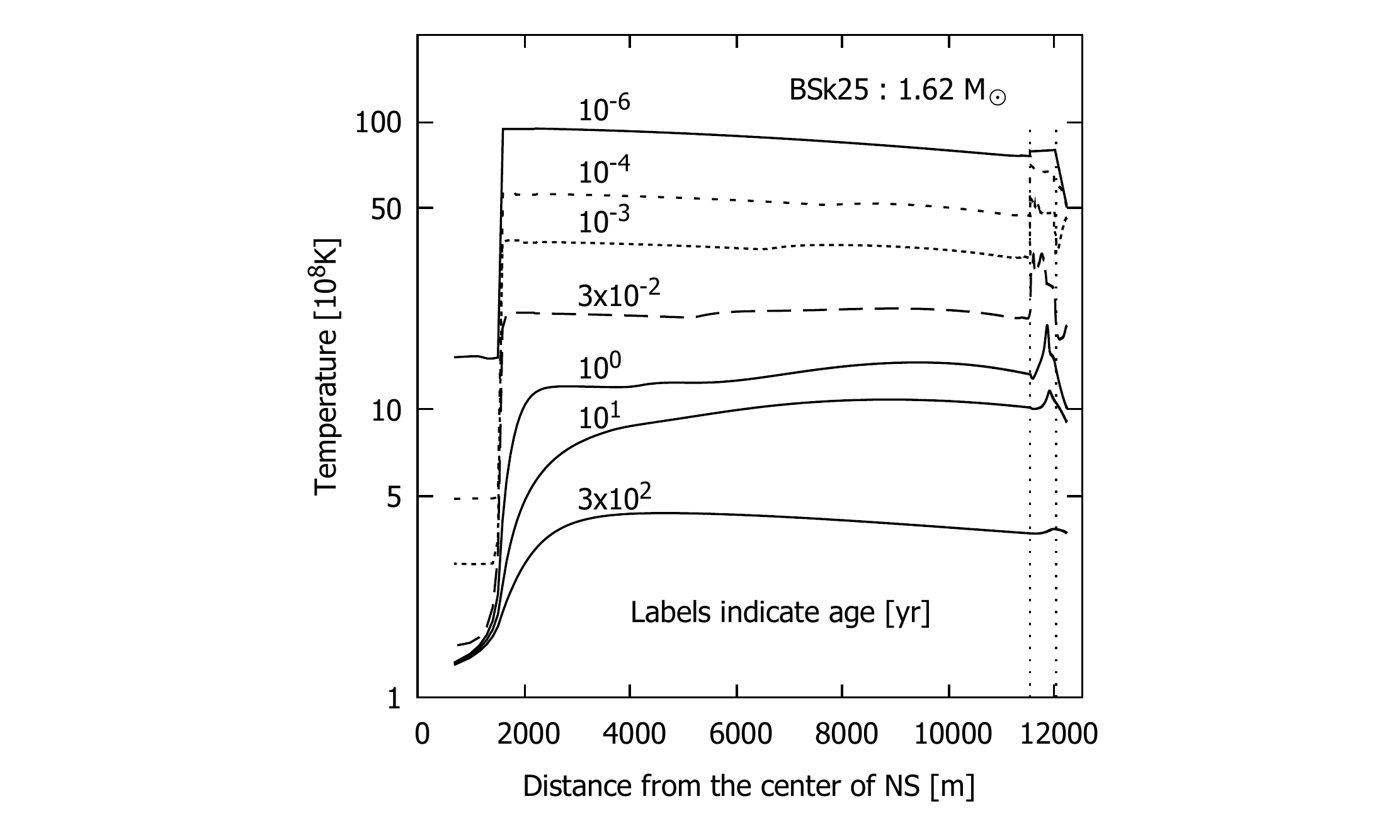}
\caption{Temperature distribution along the NS radius at different points in time, 
marked next to the curves. Vertical dotted lines denote the
boundaries between the core, inner crust, and outer crust of
the NS.}
\label{fig:temp}
\end{figure}

For a better understanding of the hybrid cooling scenario, it is instructive to 
consider the temperature evolution inside the NS, which is shown in 
Fig. \ref{fig:temp}. 
It can be seen that in a small central volume with a radius of $\sim R_{dU}$ the temperature drops sharply and after $\sim 10^{-4}$ years is below the critical value for the triplet Cooper pairing of neutrons inside this volume. This causes the activation of 
PBF neutrino emission, but not strongly, because the volume of superfluid 
neutrons at this time is small. A small jump in the PBF energy losses occurs at 
the age $4.91\times 10^{-1}$ years owing to the expansion of the relatively cold
region into a larger volume.  

The luminosity of direct Urca reactions from the small volume is also not strong. The modified Urca processes dominate in the neutrino losses in the entire volume of the NS core until its complete cooling and the onset of neutron superfluidity, which occurs approximately after $10^2$ years. 
At this time, the PBF neutrino emission from superfluid neutrons sharply increases and becomes dominant, causing rapid cooling, shown in Fig. \ref{fig:25}. 
\begin{figure}
\includegraphics[height=8.6cm,trim=130 0 130 0,clip,width=8.6cm]{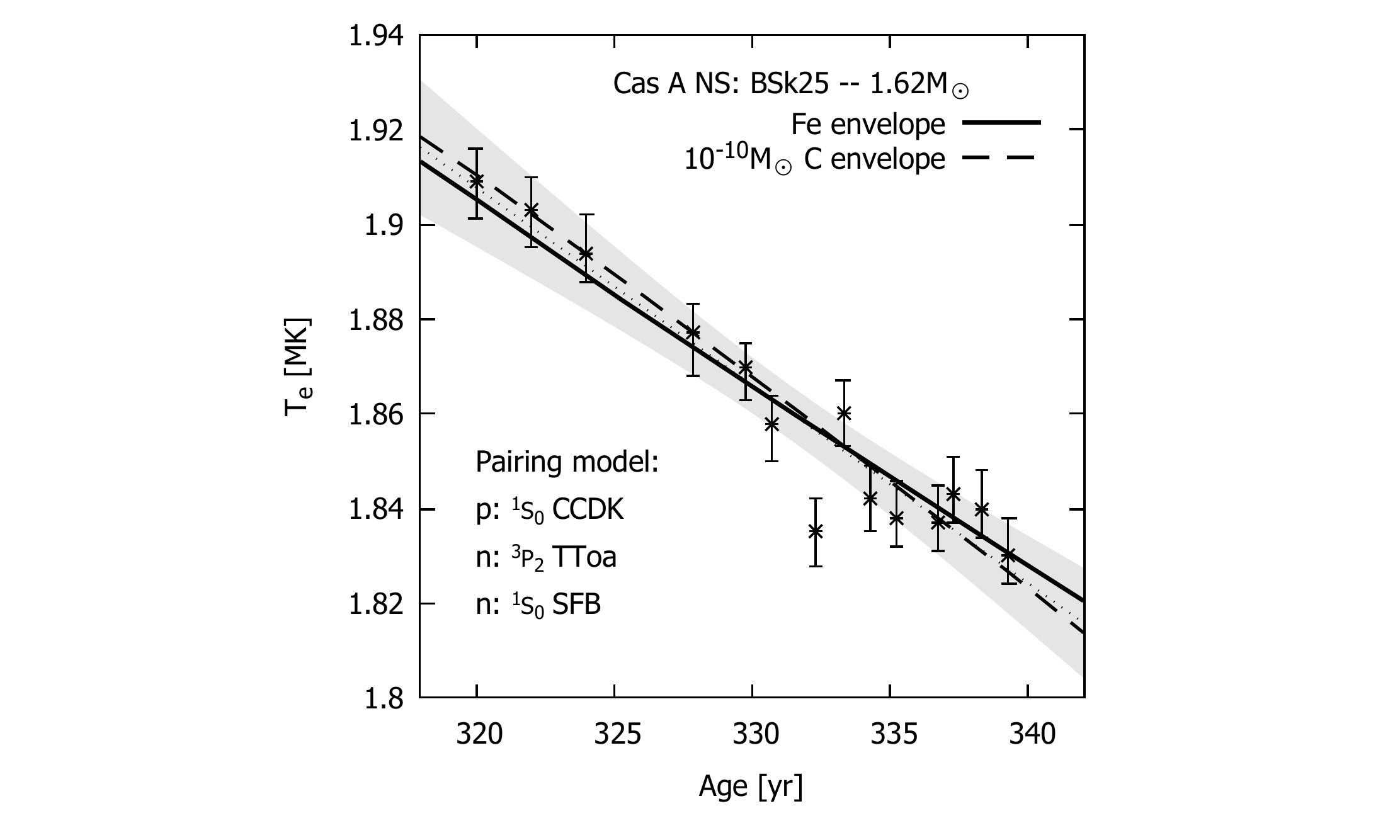}
\caption{The best fit cooling trajectories for NS with a mass 
$M=1.62M_{\odot}$ and a radius $R=12.36$ km, with iron and carbon envelopes 
in comparison with data points from \citet{Ho:2021hwy}.}
\label{fig:10}
\end{figure} 
\begin{figure}
\includegraphics[height=8.6cm,trim=130 0 140 0,clip,width=8.6cm]{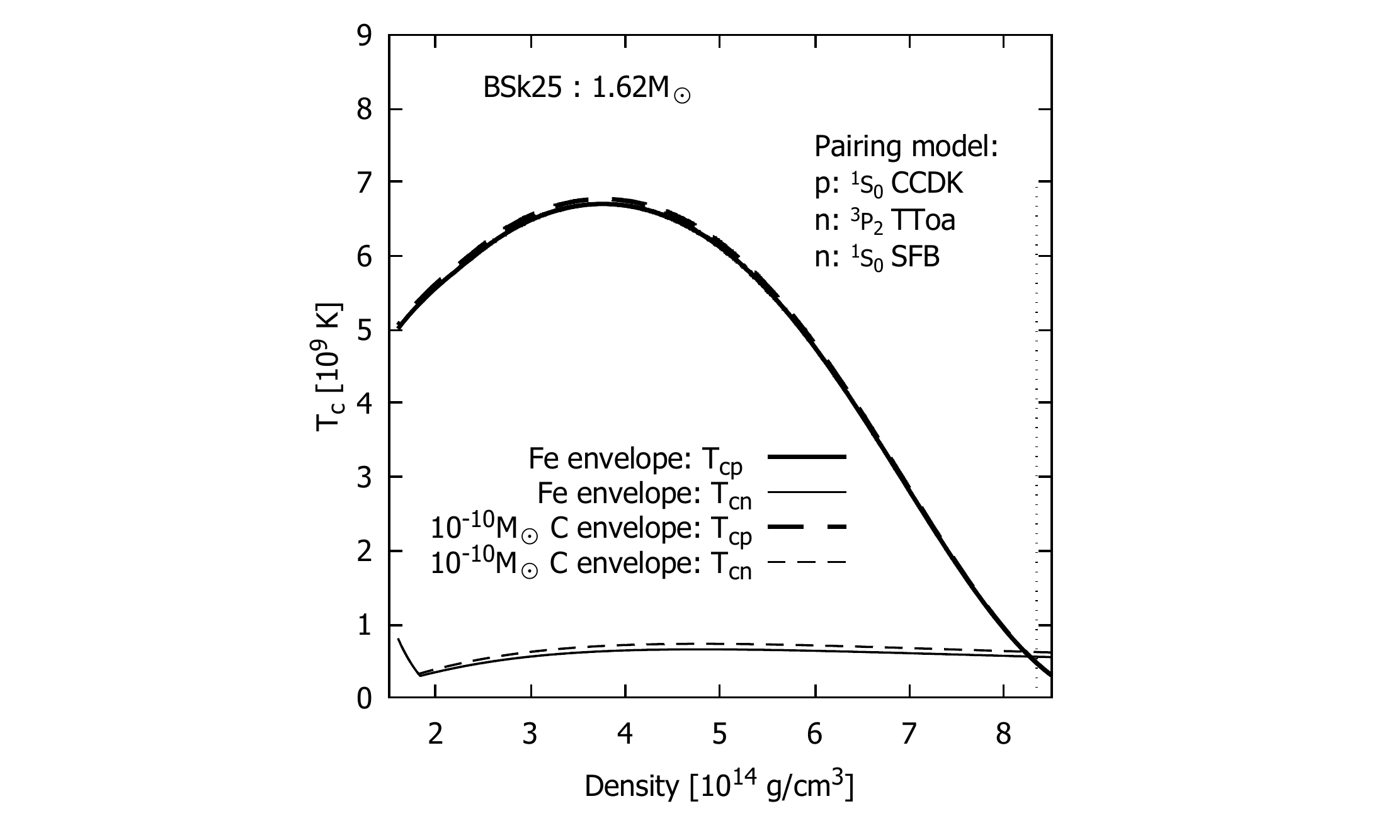}
\caption{Critical temperature $T_c$ for proton and neutron superfluidity as a 
function of mass density of a NS constructed using BSk25 EOS. Vertical 
dotted line denote the core central density of the NS.}
\label{fig:tc}
\end{figure} 

The thermal soft X-ray spectrum of Cas A is in good agreement with the model 
of the carbon atmosphere \citep{Page:2010aw,Shternin:2010qi}. 
Light elements have a higher thermal conductivity and make the envelope more 
heat-transparent \citep{2003ApJ...594..404P}. Therefore, in addition to the 
NS model with an iron envelope, the model with a carbon envelope is of interest.
According to the analysis of \citet{Shternin:2014fea}, a large 
amount of carbon in the envelope cannot be reconciled with the Cas A NS 
observations. Hence I have considered the case of a carbon layer with mass 
$\Delta M=10^{-10}M_{\odot}$ that extends down from the atmosphere to the 
bottom of the NS envelope at $\rho =10^{10}$ g cm$^{-3}$. To relate the temperature  
$T_b$ at the bottom of the heat blanketing envelope to its surface temperature $T_e$, 
which is available for observations,  I use the approximation for the $T_e-T_b$ 
dependence in a carbon-iron envelope from \citep{2016MNRAS.459.1569B}.
The best fit cooling trajectory for this case is shown in Fig. \ref{fig:10} 
together with the cooling curve for the NS with iron envelope. Excellent agreement 
with observational data in this case is achieved by slightly varying the critical 
temperatures for neutron and proton superfluidity, as shown in 
Fig. \ref{fig:tc}. 

\begin{figure}
\begin{multicols}{2}
\includegraphics[height=8.6cm,trim=120 0 130 0,clip,width=8.6cm]{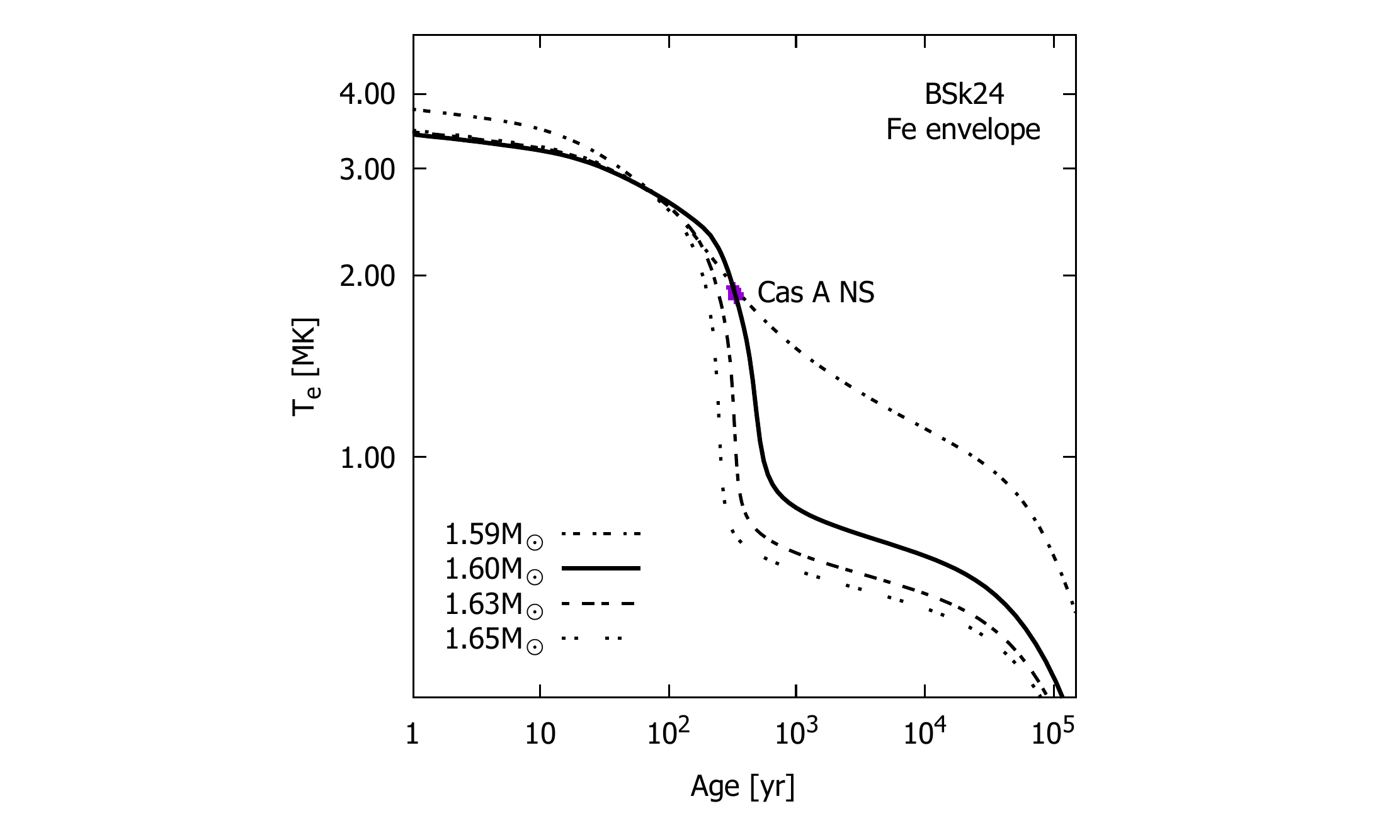}
\includegraphics[height=8.6cm,trim=120 0 130 0,clip,width=8.6cm]{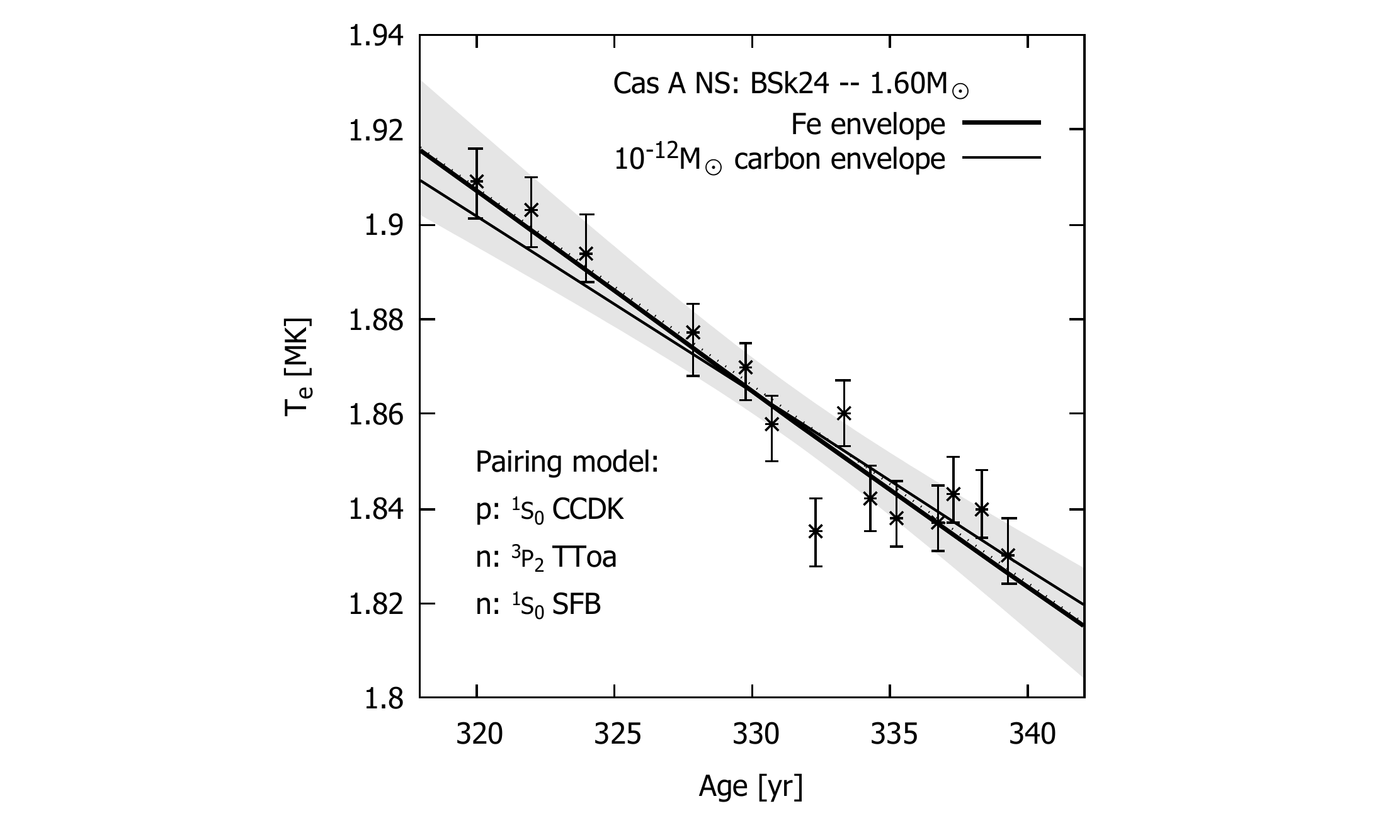}
\end{multicols}
\caption{The best fit cooling trajectories for NS with a mass 
$M=1.60M_{\odot}$ and a radius $R=12.55$ km, with iron and carbon envelopes 
in comparison with data points from \citet{Ho:2021hwy}}
\label{fig:teff24}
\end{figure}

Let us now consider the case of a NS  built from BSk24 EOS. In this case, the density threshold for direct Urca processes is $n_{dU}=0.453$ fm$^{-3}$, which corresponds to $M_{dU}=1.595M_{\odot}$ \citep{Pearson:2018tkr}, therefore, a NS with a mass $M=1.60M_{\odot}$ is the most suitable candidate for simulating the cooling of the Cas A  NS. In such a star, the nucleon matter, where the direct Urca reactions are active, 
is concentrated in a central volume with a radius
$R_{dU}=1.415\times 10^{5}$ m and have a very small mass 
$\Delta M_{dU}=4.993\times 10^{-3 }M_{\odot }$. Thus, the cooling of the NS 
occurs according to the hybrid scenario.  
The cooling trajectory of such a star with iron and carbon 
envelopes is shown in Fig. \ref{fig:teff24}. Excellent agreement with the 
observational data of Cas A NS in this case is also achieved with the aid of a small 
change in the critical temperatures for superfluidity of neutrons and protons.
\section{Discussion and conclusion}
\label{sec:disc}
I propose a natural explanation for the observed stable rapid decrease in 
the surface temperature of the young Cas A NS, using the correct form of 
neutrino energy losses from the core of the superfluid NS, which is a 
well-known problem in the paradigm of minimal cooling. I draw the reader's 
attention to the fact that the most recent EOSs: BSk24, BSk25 
\citep{Pearson:2018tkr}, which have been accurately fitted to almost all 
available atomic mass data and are constrained to fit, down to the densities 
prevailing in the cores of neutron stars predict the activation threshold 
for direct Urca processes for NS with masses close to the most appropriate 
mass of the Cas A NS, following from observations. 
This fact suggests that the evolution of the Cas A NS does not fit into the minimal 
cooling paradigm.  

To improve the Cas A NS cooling model I choose the NS mass so that the direct 
Urca processes are activated in a very small central volume of its core. 
In this case, a stable anomalous decrease in the NS surface temperature is provided by the combined losses of neutrinos from direct Urca reactions and the PBF neutrino emission that dominates in the current period.  

As a reminder, the idea of the Cas A NS  cooling owing to the direct Urca processes 
has already been expressed in \citep{Taranto:2015ubs}. Here, a scale factor 
has been used to 
significantly reduce the overall thermal conductivity to prevent cooling too 
quickly. The validity of this approach is highly questionable, since the 
lepton and nucleon thermal conductivities are known with good accuracy  
\citep[see][]{2007PhRvD..75j3004S,2012PhRvC..85b2802B,2013PhRvC..88f5805B}. 

On the contrary, the hybrid scenario of the Cas A NS cooling that I propose in this paper does not contain any adjustable parameters, except for a careful choice of the mass of NS and the generally accepted reasonable change in the critical temperature of nucleon pairing caused by the scatter in the results of numerous model calculations of the superfluid energy gap. 
This scenario unambiguously connects the used EOS and the mass of the NS, providing 
a new tool for processing observational data. 
The most suitable cooling trajectories for Cas A NS give the mass 
$M=1.62M_{\odot}$ and the radius $R=12.36$ km using EOS BSk25 or 
$M=1.60M_{\odot}$ and $ R =12.55$ km using EOS BSk24. 
The still existing large observational and theoretical uncertainties do not allow 
excluding other EOS. I have only shown that it is possible to accurately measure 
the mass of the NS using the method described. 
\section*{Acknowledgements} 
I thank Peter S. Shternin for helpful discussion of the Chandra data used here. 
\section*{Data Availability}
The data underlying this article will be shared on reasonable request
to the corresponding author.

\end{document}